\documentclass{aa}
\input{psfig.sty}
\usepackage{graphicx}   
\begin{document}
\title{The extended hard X-ray emission from the Vela Plerion}

\author{ V. Mangano\inst{1}, E. Massaro\inst{2}, F. Bocchino\inst{3}, 
T. Mineo\inst{1}, G. Cusumano\inst{1}}
\institute{
Istituto di Astrofisica Spaziale e Fisica Cosmica - Sezione di Palermo,
IASF-CNR, Via Ugo La Malfa 153, I-90146 Palermo, Italy \and
Dipartimento di Fisica, Universit\`a La Sapienza, Piazzale
A. Moro 2, I-00185 Roma, Italy \and
INAF-Osservatorio Astronomico di Palermo, Piazza dei Normanni 1, I-90100 
Palermo,
Italy 
}

\offprints{Vanessa Mangano: vanessa@pa.iasf.cnr.it}

\date{Received:.; accepted:.}

\titlerunning{The hard X-ray emission from the Vela Plerion}

\authorrunning{V. Mangano et al.}

\abstract{We present the results of a broad band (3-200 keV) 
spectral analysis of {\it Beppo}SAX and XMM-{\it Newton} observations of 
the Vela plerion. 
The hard X-ray ($>$15 keV) emission is found to be extended over a 
region of 12$'-15'$ radius, corresponding to a size of about
1.0--1.3 pc for a source distance of 290 pc. 
A single power law does not give an acceptable fit while
a broken power law or a log-parabolic law nicely fit the data. 
The former spectral model has the photon index 
$\Gamma_1$ = 1.66$\pm$0.01 for energies lower than the break value equal to 
12.5 keV and the index $\Gamma_2$ = 2.01$\pm$0.05 up to 
about 200 keV. The total X-ray luminosity of the Vela plerion
is $L_X=5.5\times10^{33}$ erg s$^{-1}$, which implies
a conversion factor of the spin--down power of $\sim$10$^{-3}$.
}
\maketitle

\section{Introduction}

The structure and evolution of filled-centre SNRs, also known as
{\it plerions} according the original definition by Weiler \&
Panagia (1978), is a rich subject that involves many physical
processes from hydrodynamics to high energy particle acceleration
and radiation.
Plerions are powered by young pulsars which inject in the remnant
a relativistic wind containing high energy particles and a magnetic
field. The physical mechanisms that convert the original 
Pointing--dominated energy flux into particle energy is not completely
understood. Furthermore, there is evidence that particles are
accelerated inside the plerion at a relatively large distance from
the pulsar but the basic mechanism is not entirely clear.
Plerion emission is observed across the entire electromagnetic
spectrum, from radio waves to the most energetic $\gamma$-rays.
However, in some frequency ranges, only little data are
available: for example, in the hard X/soft$\gamma$-rays very
little is known about the spectra of plerions.

\begin{figure*}[ht]
\centering
\includegraphics[angle=0,width=18cm]{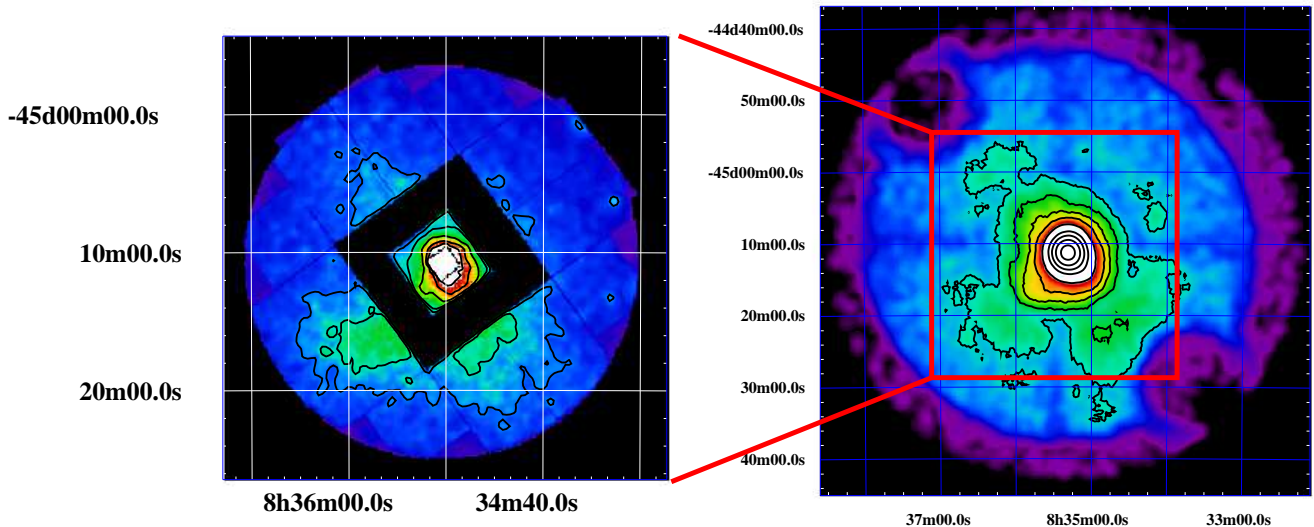}
\caption{
Left panel:  XMM--MOS1 image of the Vela PWN 
in the 3--10\,keV band;
contour levels are at 5$\times$10$^{-4}$, 1$\times$10$^{-3}$, 2$\times$
10$^{-3}$ and 8$\times$10$^{-3}$ times the brightest point. 
Right panel: MECS image of the source; 
outer contour levels are at 3$\times$10$^{-3}$, 7$\times$10$^{-3}$, 13$\times$
20$^{-3}$ and 33$\times$10$^{-3}$ times the brightest point.
Note the more extended emission in the south-west and south-east regions.
}
\label{VHf1}
\end{figure*}

The Vela SNR is one of the nearest plerions and its structure can
be studied in detail.
Its distance is estimated to be about 300 pc (Caraveo et al. 2001, 
Dodson et al. 2003), and therefore an angular size of 1$'$ 
corresponds to 0.09 pc (2.7$\times$10$^{17}$ cm).
The best {\it Chandra} X-ray images show a compact nebula with
a jet-counter jet originating from the pulsar and a double bow structure
(Helfand et al. 2001, Pavlov et al. 2001a). Kargaltsev \& Pavlov (2003)
presented maps of the 2--10 keV photon index
($F(E_\gamma)\propto E_{\gamma}^{-\Gamma}$)
with resolutions of 2$''$.5 and 10$''$ which show that the spectrum is
harder ($\Gamma\simeq1.2-1.4$) in the near surroundings of the pulsar 
and becomes softer ($\Gamma\simeq1.6-1.8$) at distances of 
$2^{\prime}-3^{\prime}$. 
A more detailed study of the inner Vela Pulsar Wind Nebula (PWN), based on 13 X-ray images,
spread over a time interval of 2.5 years, has been performed by
Pavlov et al. (2003). In particular, these authors report the
presence of bright blobs moving away along the outer jet and
fading on time scales of weeks. Photon indices in the band 
1.0--8.0 keV derived for the various structures are generally
between 1.2 and 1.5. 

Results of a spectral analysis of XMM data have been presented
by Mori et al. (2004): their spectral analysis of PN data, extracted 
from a circular region around the pulsar position within a radius of 
1$'$, gave a photon index of 1.64$\pm$0.08.

The present knowledge about the hard X-ray emission from the Vela
SNR is rather poor. The most recent data are those obtained in the
energy range 60--400 keV with OSSE-CGRO
(FOV 3$^{\circ}$.8$\times$11$^{\circ}$.4) (De Jager, Harding \&
Strickman 1996): a single power law best fit indicated a quite
hard spectrum with the rather uncertain photon index of 1.6$\pm$0.5.
It was found consistent with the previous (2--25 keV) data
obtained with the Birmingham Spacelab 2 coded mask detector
(Willmore et al. 1992), that gave $\Gamma=1.74\pm0.08$ inside
a circle of radius 6$'-$7$'$. Furthermore these authors found evidence 
for a more extended emission with a much softer spectrum. 

On the basis of a new model for the synchrotron compact nebular emission,  
Sefako and De Jager (2003) estimated for the Vela plerion the acceptable
ranges of the two main PWN parameters: the wind magnetization $\sigma$ 
(Kennel \& Coroniti 1984), whose value is expected in the range 0.05--0.5, 
and the pair production multiplicity $M$, in the interval between 300 and
700. These authors used the power law spectrum derived from OSSE data and 
assumed that the Spectral Energy Distribution (SED) peaks in the MeV range.

In this paper we present the results of the analysis of {\it Beppo}SAX data of 
the Vela Plerion  in the energy range 3--200 keV combined with the imaging and
spectral capabilities of XMM-{\it Newton}.
We derived information on the size of the source emission 
above 10 keV and {\rm obtained} an accurate evaluation of the spectral 
parameters, useful to measure the total energy output from this plerion
and for a more accurate comparison  with models
of nebular emission.

\section{BeppoSAX and XMM--Newton observations and data reduction}

The XMM-{\it Newton} observations were performed on 1 and 2 December 2000 
with an exposure of 38 and  57 ks, respectively.
We reduced the EPIC data with the XMM Science Analysis System
(SAS) version 6.0 and performed standard screening of the EPIC data.
The 2 December 2000 observation suffered high background contamination
and required a careful rejection of time intervals with soft proton flares; 
to asses the degree of soft proton contamination we used the diagnostic
defined by De Luca \& Molendi (2004).
Because of this additional screening the effective exposure in the 
2 December 2000 observation was reduced to 47 ks.
In both observations MOS1 was operated in Large Window mode, 
where the inner active region of the MOS array is the central square of
5$'$.5 side (see Fig. 1, left panel).
MOS2 was working in Small Window mode with an active
central square of 1$'$.8 side only and was not included
in the analysis of the wider central plerion emission.
PN was working in Small Window mode but the target was not centred 
in the active square region   and  therefore    
PN data were not included in our analysis.

Diffuse and asymmetric emission is visible in the image 
of Fig. 1 (left panel) beyond 8$'$ mostly in South-East
and South-West directions.
Comparing the X-ray contour with the radio images recently
obtained by Dodson et al. (2003) we note that the structure of the 
diffuse emission is correlated with that observed at radio frequencies.
{\rm We extracted spectra from circular regions centred on the
source with a radius of  0$'$.5, 1$'$.0, 1$'$.5 and 2$'$.0
(therefore including the central source)
and, to investigate better the spectral changes moving outward
in the nebula, also from annular regions centred at the source
with radii 0$'$.5--1$'$.0, 1$'$.0--1$'$.5, 1$'$.5--2$'$.0 and 
8$'$--12$'$ (outside the gap).}
We also extracted spectra from two circles of 2$'$ radius
in the South-East (RA: 08h35m52s, DEC: -45d16m26s) and
South-West (RA: 08h34m58s, DEC: -45d17m04s) regions where,
according to Fig. 1, diffuse emission is visible.
Spectra and effective area corresponding to the two observations
were summed using the tasks {\it mathpha} and {\it addarf} from the
HEASOFT 5.3.1 package.

The spectrum of the background for the MOS1 was extracted
from a set of 10 public galactic plane observations
at $l$=310$^{\circ}$--312$^{\circ}$ 
from which we have removed the detected sources,
and, for comparison,
from {\it closed} datasets (Marty et al. 2002) containing only 
the instrumental and particle components of the background.
Results obtained using these two different background spectra 
were always compatible.
Vignetting correction was properly applied to both 
source and background spectra.
Spectral response matrices and effective area were calculated
for each region via the SAS tasks {\it rmfgen} and 
{\it arfgen}, respectively, according to the standard recipe for
vignetting--corrected extended sources.

The {\it Beppo}SAX observation of the Vela pulsar region was
performed on 18--20 November 1997 for a total net exposure of 84
ks in the MECS and 38 ks in the PDS (for details on the {\it Beppo}SAX
instrumentation see Boella et al. 1997). Standard procedures and
selection criteria were applied to the data using the SAXDAS
v.2.0.0 package. 
The image extracted from the MECS (Fig. 1, right panel) 
presents regions of diffuse emission in the South-East 
and South-West directions just as in the XMM image.
The extended structure of the source requires that a
suitable auxiliary matrix must be created to correct for the
redistribution due to the instrument point spread function and for
the vignetting  relevant at distances larger than about $4'$
from the centre of the FOV (Fiore et al. 1999). 
A proper response matrix can be generated for the MECS data 
using the SAXDAS {\it effarea} command (for details see Molendi 1998, D'Acri
et al. 1998) and the average radial profile of the Vela plerion derived
from the XMM-{\it Newton} image interpolated with a spline function in the gap
region and extrapolated with a constant up to 15$'$.
We produced ad-hoc matrices for the circular collecting areas of radii
equal to  4$'$, 8$'$, 12$'$ and 15$'$. 
{\rm The poor angular resolution of the MECS and the difficulty
of producing a good response matrix for annular regions did not allow
us to perform a spatial differential study of the X-ray emission, 
as done for MOS data.}

The spectrum of the background for the MECS was extracted from
archive blank fields. 

PDS data were collected using the standard rocking observation mode
with two detectors pointing at the source and the other two in
offset directions to measure the background simultaneously.
The net count rate in the whole energy band was 1.7 c/s.
The PDS field of view, 1.3$^{\circ}$ FWHM, is large enough to include
the entire plerion emission. 
It is, however, much narrower than that of OSSE ($3^{\circ}.8\times11^{\circ}.4$ 
FWHM).
We also checked that PDS field of view does not contain bright sources
that could contaminate the Vela  spectrum.

Events with an energy higher than 3 keV were selected for the analysis
to exclude the soft thermal contribution from the pulsar 
(Pavlov et al. 2001b) and from the Vela shell (Bocchino et al. 1999).
Effective ranges were  3--10 keV for MECS and MOS and 15--200 keV
for PDS. For the same reason we did not consider LECS data 
in the analysis. 
Spectral fits were first evaluated separately for the MOS, MECS
and PDS data and after, combined fits of MOS--PDS and MECS--PDS
data sets were considered. 

A relevant point we stress when working with spectra
obtained with several instruments is that a  proper evaluation of the
inter-calibration factors is required.
For the {\it Beppo}SAX NFIs, the accurate ground
and in-flight calibrations were used to establish that the
admissible ranges for the factor between MECS  and PDS for point
sources ($f_{MP}$) is 0.77 $\leq f_{MP}\leq$0.93, reduced to
0.86$\pm$0.03  for sources with a PDS count rate higher than 2 ct/s
(Fiore et al. 1999).
For extended sources, as for the Vela plerion, we expect that the best
fit estimate of $f_{MP}$ would result in agreement with the above value
only when the  emission observed by the PDS is fully selected in the
MECS image.
Moreover, MOS--PDS intercalibration factor is expected in a similar range
because the intercalibration analysis between MOS and MECS shows a
good agreement between the two instruments (Kirsh 2004).
Values of the intercalibration factors well above the expected range
can be an indication that MECS flux, for a given collecting
area, includes only a fraction of the signal detected by the PDS. 
Thus it is possible to use this parameter to obtain information {\rm on}
the angular size of the source responsible for the hard X-ray ($>$15 keV)
emission.

The uncertainties reported in the following are at 1 standard
deviation for one interesting parameter.

\section{Spectral analysis}

\subsection{Single instrument spectral analysis}

Spectra of each instrument were fitted  with a single power-law model.
A plot of the  MOS and MECS photon index as a function of the
extraction radius for all the {\rm circular} regions centred at the source 
is shown in Fig. 2.
The value of $\Gamma$ changes with the extraction radius:
for a radius $\leq$2$'$ it is around 1.55, while for higher
radii it increases to  1.67, in agreement, within the errors, with the value
reported by Willmore et al. (1992), indicating a quite mild softening
of the spectrum in the outer region of the PWN.
Figure 3 shows the fluxes in the 3-10 keV band as a function
of the spectra extraction radius.
The flux  increase  within 4$'$ quantifies the amount of emission
originating from the inner  region. 

{\rm The spatially differential spectral analysis confirms that
the inner region of the plerion has a flatter spectrum than
the outer one. The average spectral index of the three annuli
is 1.60$\pm$0.02 (the third annulus differs from this value
only by two standard deviations) to be compared with 1.90
found at distances greater than 8$'$. The small difference 
(at about two standard deviations) between the mean index of the 
three inner annuli and that found for the circular region of
0$'$.5 radius is likely due to the contribution of the
inner part of the PWN
which is about 40\% of the total as derived 
from the 3--10 keV flux ratio.}

{\rm The spectral index variation in the inner three MOS annuli 
is in agreement with the Chandra spectral index map
presented by Kargaltsev \& Pavlov (2003).}

Fitting the PDS spectrum in the energy range 15$-$200 keV gives a
photon index equal to 2.00$\pm$0.05 with the $\chi^2_r$ of 0.73 for
16 d.o.f. and a flux of (15.7$\pm$0.4)$\times$10$^{-11}$ erg cm$^{-2}$ s$^{-1}$.
This photon index, significantly higher than those found in the lower
energy range, indicates that a spectral steepening occurs at higher energies.
Moreover, it is  higher than the value obtained
with OSSE (De Jager et al. 1996), although statistically compatible
because of the large uncertainty in their estimate.
A summary of the best fit results for the various extraction radii
of the imaging instruments and PDS are given in Table 1.

\begin{table*}[ht]
\centering
\caption{Best fit parameters for a single power law of MOS and MECS. The values given in the
first section are for the spatially integrated analysis, while those of the second section are
for the MOS annular regions.}
\begin{tabular}{cccccc}
\hline
     & Radius & $\Gamma^{\mathrm{(a)}}$ & $K^{\mathrm{(a,b)}}$ &$F^{\mathrm{(a,c)}}$ & $\chi^2_r$ (d.o.f.) \\
\hline
\hline
MOS  & ~0$'$.5      & ~~1.50$\pm$0.02   & ~~{\rm 0.39$\pm$0.01}  & ~~{\rm 1.785}$\pm$0.009 & ~~1.16 (253) \\
MOS  & ~1$'$~~      & ~~1.55$\pm$0.01   & ~~0.95$\pm$0.02        & ~~{\rm 3.98}$\pm$0.01   & ~~1.08 (324) \\ 
MOS  & ~1$'$.5      & ~~1.56$\pm$0.01   & ~~1.09$\pm$0.02        & ~~{\rm 4.49}$\pm$0.01   & ~~1.11 (331) \\
MOS  & ~2$'$~~      & ~~1.55$\pm$0.01   & ~~1.18$\pm$0.02        & ~~{\rm 4.95}$\pm$0.01   & ~~1.07 (337) \\
\hline 
MECS & ~4$'$~~      & ~~1.61$\pm$0.02   & ~~1.50$\pm$0.04        & ~~{\rm 5.66 $\pm$0.03} & ~~0.96 (~66)  \\
MECS & ~8$'$~~      & ~~1.67$\pm$0.02   & ~~1.88$\pm$0.05        & ~~{\rm 6.39}$\pm$0.03   & ~~0.93 (~66)  \\
MECS & 12$'$~~      & ~~1.66$\pm$0.02   & ~~2.17$\pm$0.05        & ~~{\rm 7.50}$\pm$0.03   & ~~1.02 (~66)  \\
MECS & 15$'$~       & ~~1.66$\pm$0.01   & ~~2.62$\pm$0.09        & ~~{\rm 9.06}$\pm$0.04   & ~~1.02 (~66)  \\
\hline
PDS  & FOV          & ~~2.00$\pm$0.05   & ~~{\rm 6.19}$\pm$0.95$^{\mathrm{(d)}}$ & ~~{\rm 15.9}$\pm$0.4$^{\mathrm{(e)}}$  & ~~0.73 (~16) \\
\hline
\hline
{\rm MOS  }&{\rm ~0$'$.5--1$'$ }&{\rm ~~1.60$\pm$0.02  }&{\rm ~~0.56$\pm$0.02  }&{\rm ~~2.150$\pm$0.009 }&{\rm ~~1.05 (209) }\\
{\rm MOS  }&{\rm ~1$'$--1$'$.5 }&{\rm ~~1.61$\pm$0.03  }&{\rm ~~0.14$\pm$0.01  }&{\rm ~~0.528$\pm$0.005 }&{\rm ~~0.88 (~72) }\\ 
{\rm MOS  }&{\rm ~1$'$.5--2$'$ }&{\rm ~~1.50$\pm$0.05  }&{\rm ~~0.08$\pm$0.01  }&{\rm ~~0.361$\pm$0.004 }&{\rm ~~0.77 (~49) }\\
{\rm MOS  }&{\rm ~8$'$--12$'$  }&{\rm ~~1.90$\pm$0.06  }&{\rm ~~0.51$\pm$0.05  }&{\rm ~~1.165$\pm$0.02  }&{\rm ~~1.13 (231) }\\   
\hline
\hline              
\end{tabular}
\begin{list}{}{}
\item[$^{\mathrm{(a)}}$] Errors correspond to 1 standard deviation for one interesting parameter. 
\item[$^{\mathrm{(b)}}$] Power law normalisation $K$ is expressed in units of 
        $10^{-2}$ photons cm$^{-2}$ s$^{-1}$ keV$^{-1}$ at 1 keV.
\item[$^{\mathrm{(c)}}$] Flux is in units of $10^{-11}$erg cm$^{-2}$ s$^{-1}$. 
        It is relative to the 3-10\,keV band for the MOS/MECS and to the
        20--100\,keV band for the PDS. 
        {\rm The reported errors are statistical. Systematical errors amount 
             to 5\% for MOS flux values and 10\% for the MECS.} 
\item[$^{\mathrm{(d)}}$] {\rm The PDS normalisation is obtained dividing the 
        single power law fit result (5.2$\pm$0.8) 
        by the intercalibration factor $f_{MP}=0.84$ obtained 
        in the combined fit with the MECS spectrum extracted 
        from the circle of 15$'$ radius.}
\item[$^{\mathrm{(e)}}$] The PDS flux value 
        {\rm is that obtained from the single power law fit}
        ({\rm 13.39$\pm$0.34}) {\rm divided} by the intercalibration factor $f_{MP}=0.84$.
\end{list}
\end{table*}

\begin{figure}
\centering
\includegraphics[angle=-90,width=8.7cm]{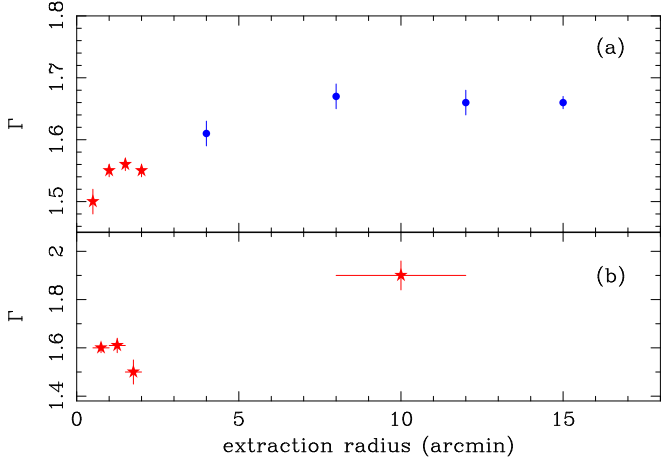}
\caption{ {\rm Panel (a):} 
the power law photon index in the 3$-$10 keV band of MOS
(stars) and MECS (filled circles) spectra at varying extraction radius.
{\rm Panel (b): the power law photon indices in the 3$-$10 keV band of MOS 
spectra extracted from the annuli listed in Table 1.}}
\label{VHf2}
\end{figure}

\begin{figure}
\centering
\includegraphics[angle=-90,width=8.7cm]{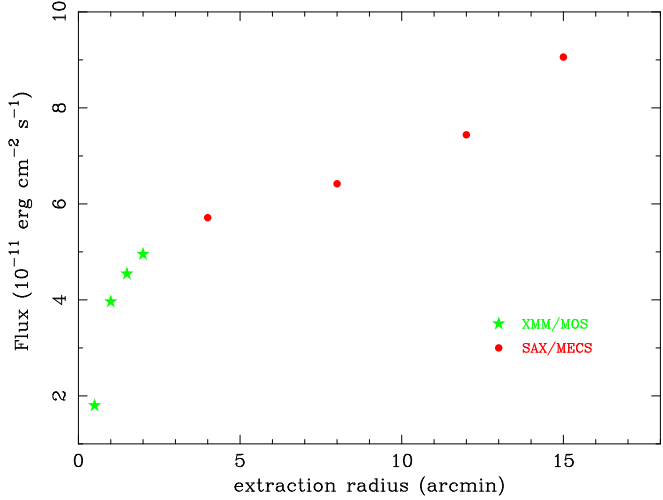}
\caption{Vela PWN flux in the 3--10\,keV band at varying extraction
radius.
}
\label{VHf3}
\end{figure}

We also investigated the spectral properties in the outer part 
of the plerion. The XMM spectra extracted from two circular regions
having a diameter of 4$'$ in the South-East and South-West 
were also fitted with a simple power-law model. 
We obtained
$\Gamma=1.9\pm0.1$, $K=(5.0\pm1.7)\times 10^{-4}$ photons cm$^{-2}$ s$^{-1}$ keV$^{-1}$ 
($\chi^2_r=0.94$, 181~d.o.f) in the SE region and 
$\Gamma=1.9\pm0.1$, $K=(5.5\pm1.5)\times 10^{-4}$ photons cm$^{-2}$ s$^{-1}$ keV$^{-1}$ 
($\chi^2_r=1.01$, 186~d.o.f) in the SW region, respectively.
The 3--10\,keV band fluxes per unit solid angle from the two regions
were $(8.4\pm0.3)\times 10^{-7}$ erg cm$^{-2}$ s$^{-1}$ sr$^{-1}$
and $(9.6\pm0.5)\times 10^{-7}$ erg cm$^{-2}$ s$^{-1}$ sr$^{-1}$.

\begin{figure}[ht]
\centering
\includegraphics[angle=-90,width=8.7cm]{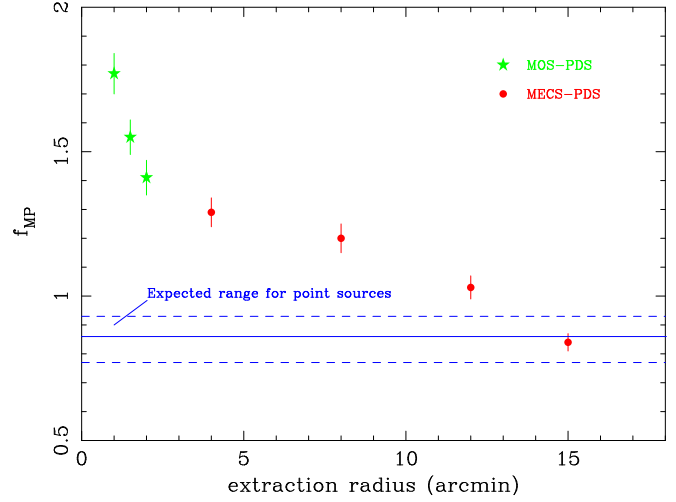}
\caption{Best fit values of $f_{MP}$ in the combined log-parabolic fit 
with the PDS at varying extraction radius.
}
\label{VHf4}
\end{figure}

\begin{figure}[hb]
\centering
\includegraphics[angle=-90,width=8.7cm]{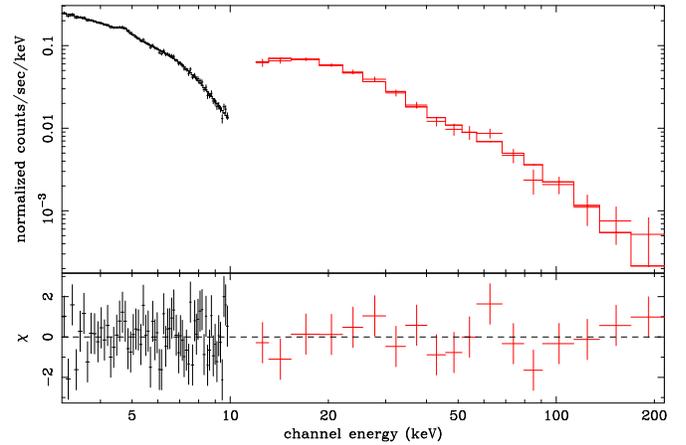}
\caption{Best fit of MECS and PDS data with a broken 
power law model ($15^{\prime}$ extraction radius for the MECS data).
See Table 2 for the best fit parameter values.
}
\label{VHf5}
\end{figure}

\subsection{Combined instrument spectral analysis}

The finding that the 15--200\,keV X-ray photon index is significantly larger
than those measured below 10 keV indicates that a single power law
is not able to fit the broad band spectrum of the Vela PWN.
We therefore performed a spectral analysis combining the PDS data 
with those of MOS/MECS.
The single power-law model was rejected because the 
$\chi^2_r$ values of the combined fits were in the range
1.34(341 d.o.f.) -- 1.84(83 d.o.f.). Moreover, PDS residuals show 
very large deviations with $\chi^2_r$ always higher than 4 (15 d.o.f). 

Two other models were then considered:
a broken power law:
\begin{eqnarray}
F(E) & = & K\, (E/1\,{\rm keV})^{-\Gamma_1} \hspace{1.85cm}  {\rm for}\ E < 
E_b
\nonumber
\\
F(E) & = & K E_b^{(\Gamma_2- \Gamma_1)}\, (E/1\,{\rm keV})^{- \Gamma_2} 
    \hspace{0.5cm}  {\rm for}\ E > E_b
\end{eqnarray}\
and a continuous steepening law, characterised by a linear dependence of
the spectral slope upon the logarithm of energy:
\begin{equation}
F(E) = K (E/ 1\,{\rm keV})^{-(\alpha + \beta~Log~(E/ 1\,{\rm keV}))}
\end{equation}

\noindent

The fits of the broken power law (Eq. 1) to the spectra for
different extraction radii gave acceptable $\chi^2_r$ for all regions
but showed a strong correlation between the intercalibration factor $f_{MP}$
and the break energy $E_b$. To extract reliable values from this model
it is necessary to fix one of these two parameters.

The log-parabolic law (Eq. 2) gave acceptable fits for all extraction 
radii without evidence of a correlation between the parameters.
The best fit values of $f_{MP}$ are outside the expected range with  
extraction radii lower than 15$'$ as shown in Fig. 4. 
This result indicates  that to match the PDS signal to the 3-10\,keV flux 
we must extract events in a MECS region with a radius of about 15$'$. 

To further investigate the extension of the region 
responsible for the emission above 15 keV
we fixed $f_{MP}$ to 0.93 (the highest acceptable value) and, 
fitting  the 12$'$ spectrum with Eq. (2), we obtained 
an acceptable $\chi^2_r$ of 0.98 (83 d.o.f.) 
and an acceptable $\chi^2_r$ relative to PDS data only of 1.09 (15 d.o.f).
However, the same procedure  on  the 8$'$ spectrum gave 
a marginally acceptable fit for the two datasets ($\chi^2_r=1.3$, 83 d.o.f.) 
but not acceptable for PDS data only ($\chi^2_r=2.90$, 15 d.o.f.). 
Thus, we are strongly confident that the extension 
of the hard X-ray emitting region is at least 12$'$ from the pulsar.

\begin{table}[ht]
\centering
\caption{Best--fit model parameters of the combined fit MECS-PDS 
relative to  the MECS extraction radius of 15$'$
}
\begin{tabular}{cc}
\hline
\multicolumn{2}{c}{Log-parabolic Model} \\
\hline
\hline
$\alpha$  &  1.36$\pm$0.05 \\
$\beta$    & 0.21$\pm$0.03 \\
$K$ ($\times$ 10$^{-2}$) &  2.08$\pm$0.10 \\
$f_{MP}$   &  0.84$\pm$0.03 \\
 $\chi^2_r$ (d.o.f.) & 0.76 (82) \\
\multicolumn{2}{c}{   } \\
\hline
\multicolumn{2}{c}{Broken power-law  Model} \\
\hline
\hline
$\Gamma_1$ & 1.66$\pm$0.01 \\
$\Gamma_2$ & 2.01$\pm$0.05 \\
$E_{b}$ & 12.5$\pm$1.5 \\
$K$ ($\times$ 10$^{-2}$) & 2.62$\pm$0.06 \\
$f_{MP}$  & 0.84 (fixed) \\
$\chi^2_r$ (d.o.f.) & 0.96 (82) \\
\end{tabular}
\end{table}

We also fitted the MECS spectrum extracted from a 15$'$ radius and the PDS 
spectrum with the broken power-law model, fixing the intercalibration 
factor to the value obtained in the log-parabolic fit.
The fit gave an acceptable $\chi^2_r$. The spectrum and the {\rm best fit 
model are shown in Fig. 5, and the parameters' values are given in Table 2. 
Note that the hard branch of the spectrum is fully consistent
with the fit of PDS as single instrument (see Table 1).}

We then verified if the models used to describe the spectral distribution 
of the Vela PWN in the 3-200\,keV were consistent with the EGRET upper 
limits reported by Kanbach et al. (1994) for the unpulsed component observed 
from the region of PSR\,0833-45. We extrapolated the two best fit spectra
of the combined MECS--PDS data sets (15$'$ extraction radius for the MECS) 
in the EGRET range and the results are shown in Fig. 6. 
While the log--parabolic spectrum is much lower than the $\gamma$-ray
upper limits, that derived from the broken power law is only marginally
compatible with them. It is possible, however, that the plerion spectrum
will show another break and/or a cut--off at energies in the range
between 200 keV and $\sim$10 MeV.
In the same figure we reported also the {\rm upper limits of Mignani et al. 
(1989) for the optical nebular flux which are in satisfactory agreement 
with the low frequency extrapolation of the inner plerion spectrum}.

\begin{figure*}[ht]
\centering
\vspace{1.0cm}
\includegraphics[angle=-90,width=10.5cm]{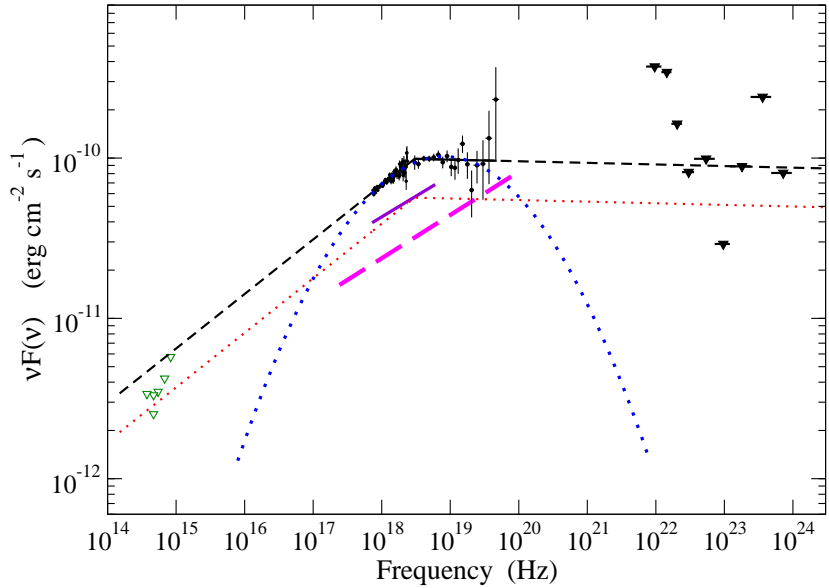}
\caption{The optical to X-ray spectral energy distribution
of the Vela PWN compared with other literature results. 
Our SEDs are the broken power law (solid and short-dashed lines) 
and the log-parabola (dotted line) to which are superposed the 
deconvolved spectral data. 
The thick solid line is the spectrum measured by the Birmingham 2
experiment (Willmore et al. 1992) and the dotted lines represent
the broken power law SED rescaled to the flux measured
by the MECS with an extraction radius of 4$'$.
The thick long dashed line is {\rm the broad band spectrum 
of the compact nebula  based on OSSE data (De Jager 1996)
and used by by Sefako and De Jager (2003). 
Optical upper limits (empty triangles) are taken from Mignani et al. (2003)
extrapolated to an area of 10$^4$ arcsec$^2$ to cover all the
inner part of the plerion.}
The triangles in the $\gamma$--ray band represent EGRET upper limits
from  Kanbach et al. (1994).
}
\label{VHf6}
\end{figure*}

\section{Discussion}

The broadband X-ray spectrum of the Vela PWN derived from
the combined analysis of {\it Beppo}SAX and XMM-{\it Newton} data, 
together with the information from high resolution images 
of Chandra (Pavlov et al. 2001a,b, 2003), provided a more
complete detailed picture of the spectrum and the structure 
of this source useful for the understanding of the main 
physical processes responsible for the energy transfer and 
radiation.

We can distinguish two main regions where X rays are emitted:
a central source of high brightness with an angular diameter of
3$'$--4$'$ surrounded by a much fainter region extending
mainly in the SE {\rm and} SW sector up to distances of about 12$'$--15$'$,
corresponding to a distance from the pulsar of about 1.2 pc.
This picture has a good correspondence to the radio images
recently obtained by Dodson et al. (2003). These authors report 
the presence of an extended and highly polarised 
non-thermal emission out to their field limit of about 6$'$ 
from the pulsar. The spatial structure is mainly elongated in
the NE-SW direction, but towards the NE it is limited by a quite sharp
boundary, not detected in the other direction. Furthermore, the
images of Dodson et al. (2003) suggest the presence of another
extended component in the SE direction at distances larger than
$\sim$ 6$'$. Likely, these two outer components are the radio
counterparts of the extended X-ray emission detected in the
MOS and MECS images. 

Our spectral analysis confirms that the central region has a harder
spectrum than the outer one, with a photon index of about 1.6, but
possibly close to 1.5 as indicated by the MOS spectra with the
smallest extraction radius. Such a hard spectrum was also observed
in the high resolution X-ray images obtained by Pavlov et al.
(2003) with the Chandra observatory.
The photon index from MOS data in the outer region is indeed close to 2.
 
The spectrum at energies higher than $\sim$15 keV is well 
fitted by a single power law with a photon index 2.
A possible interpretation of these results is that
the emission from the extended region is described by such a spectrum, 
while that of the central PWN must show a break at $\sim$12 keV
with a photon index change {\rm of} about 0.5.

{\rm This behaviour does not confirm the claim of De Jager et al. (1996)
about the spectrum of the central compact nebula  extending
unbroken up to 400 keV. }

In Fig. 6 we compare our broad band results with the hard X-ray SEDs
observed by Birmingham 2 (Willmore et al. 1992) and OSSE-CGRO experiments
(De Jager et al. 1996).
Our spectrum is well above both these data, however, the discrepancy
with the data of Willmore et al. (1992) can be explained by the 
different source extension taken into account. 
The data of the Birmingham 2 experiment were obtained modelling
the coded mask image with a central compact source plus an extended 
emission. The SED of Fig. 6 refers to the central source only and when 
scaling our SED to the flux measured by the MECS with an extraction radius
of 4$'$ (see Table 1) we found a good agreement.
At variance, the discrepancy with the spectrum derived from the OSSE-CGRO 
observation (De Jager et al. 1996) is not simply explained by the
source extension and can be due to some systematic effect possibly due
to the much wider field of view of this experiment. In particular,
when these data are extrapolated from 60 keV to energies of $\sim$1 keV
(Sefako and De Jager 2003), the source X-ray luminosity can be largely
underestimated.
We computed the isotropic X-ray luminosity of the Vela plerion by
integrating between 0.1 and 200 keV the broken power law spectrum 
taken over the largest radius and obtained $L_X=5.5\times10^{33}$ erg
s$^{-1}$. This luminosity is more than a factor of 2 higher than
that derived from the best fit OSSE spectrum in the same range.
Note that about 45\% of this luminosity is emitted at energies lower 
than $E_b$.

{\rm A comparison of the PDS flux of the Crab in the 20-200 keV range
and the corresponding OSSE flux based on Ulmer et al. (1995)
also shows a discrepancy of the order of 30\%.}

Pavlov et al. (2001a) estimated that the 1--10 keV luminosity
of the PWN within about 1$'$ from the pulsar is 6$\times$10$^{32}$ erg
s$^{-1}$, in a good agreement with that derived by us from the MOS data 
for a similar extraction radius (Fig. 3).
Such a luminosity corresponds to about 1/4 of the emission from the
whole source and implies that the outer nebula emits a relevant fraction 
of the total X-ray power.
This more precise value can be useful to improve the estimates
of the magnetization parameter $\sigma$ and the pair multiplicity $M$
of the Vela plerion. We did not develop a complex physical model of
the source for this purpose, which is beyond the scope of the present 
paper. However, one can reasonably expect that the greater luminosity
obtained by us would imply a higher number of particles and consequently 
a higher value of $M$. If confirmed by numerical models, like that
by Sefako and De Jager (2003), this finding can be the reason for the
too low value found by these authors.
  
Furthermore, we found that the X-ray emission from the plerion $L_X$ is
a factor of the order of $\sim$10$^{-3}$ of the pulsar spin-down 
luminosity $\dot{E} = 6.7\times 10^{36}$ erg s$^{-1}$, one order of 
magnitude higher than the value given by Pavlov et al. (2001a). 

It is possible to show, using a simple dimensional approach,
that the nebular extension measured in the
MECS images and used in our analysis (i.e. a radius 12$'$--15$'$,
approximately corresponding to 1.0--1.3 pc) is compatible with the radiative 
lifetime of high energy electrons $t_{sy}$ emitting hard X rays. 
According to a general plerion model we assume that electrons are
injected at the pulsar wind shock, which for Vela is at an angular 
distance from the pulsar of 53$''$ (Helfand et al. 2001), and then 
diffuse throughout the plerion.
From the synchrotron theory the radiative lifetime of 
the electrons which emit photons of energy $E_{\gamma}$ is:
\begin{equation}
t_{sy} = 2.15 \times 10^9 \bigg(\frac{E_{\gamma}}{{\rm 1 keV}}\bigg)^{1/2}~\bigg(\frac{B_{\perp}}{{\rm 10^{-4}G}}\bigg)^{-3/2}  ~~~~~~{\rm s}
\end{equation}
where $B_{\perp}$ is the average transverse component of the magnetic
field. The typical time scale to escape the nebula of 
radius $R$ (in pc units) can be written as
\begin{equation}
t_{e} \simeq R/v_d = 1.03\times 10^8 ~a~R   ~~~~~~~~{\rm s}
\end{equation}
where {\rm 1/$a$} is the outward diffusion velocity in units of the speed of 
light ($v_d=c/a$); for example, if $v_d$ is equal to the sound velocity 
in a relativistic plasma then $a=\sqrt{3}$ (De Jager et al. 1996). 
The two above equations can be used to estimate the highest energy of the
photons emitted by electrons able to reach the outer boundary of the
plerion (i.e. $t_e \simeq t_{sy}$) and then we obtain:
\begin{equation}
E_{\gamma} \simeq \frac{435}{a^2}\bigg(\frac{R}{{\rm 1pc}}\bigg)^{-2} \bigg(\frac{B_{\perp}}{{\rm 10^{-4}G}}\bigg)^{-3}  ~~~~{\rm keV}~ 
\end{equation}

The typical nebular magnetic field is estimated to be about 10$^{-4}$ G 
(Sefako \& De Jager 2003, Pavlov et al. 2003) and therefore, for any
reasonable choice of $a$, photons of energy of $\sim$ 100 keV can be 
efficiently radiated up to distances of about 1 pc from the wind 
shock region without need for a further acceleration. 
Note that these relations imply a maximum Lorentz factor for the electrons 
of $\sim$3$\times$10$^8$, as also estimated in the above quoted papers.

An interesting consequence of Eq. (5) is that photons with energy of
$\sim$ 10 MeV should be emitted from regions much closer to the wind shock
and therefore the outer nebula spectrum should show a further 
steepening at energies higher than a few hundred keV. The occurrence
of such a high energy break could be detected by the study of the
DC component of the Vela pulsar with the present (INTEGRAL) and 
the next generation of high sensitivity low-energy 
$\gamma$ ray telescopes.

\begin{acknowledgements}
We are grateful to the referee O. De Jager for his comments which improved 
our paper. We are also indebted to Bruno Sacco for useful discussions.
This work has been partially supported by INAF.
\end{acknowledgements}


\begin{thebibliography}{}

\bibitem[Bocchino et al. 1999]{bocchino99} Bocchino F., Maggio A., Sciortino S. 1999,
A\&A 342 839
\bibitem[Boella et al., 1997]{boella97a} Boella G., Butler R.C.
et al. 1997, A\&AS 122, 299
\bibitem[Caraveo et al. 2001]{Caraveo} Caraveo P., De Luca A. et al. 2001,
ApJ 561, 930
\bibitem[D'Acri et al., 1998]{d'acri} D'Acri  F., de Grandi S.,
Molendi S. 1998, in The Active X-ray Sky: Results from {\it Beppo}SAX and RXTE,
ed. L. Scarsi, H. Bradt, P. Giommi, \& F. Fiore; Nuclear Physics B, Proc.
Suppl. 69, 581
\bibitem[DeLucaMol 2004]{DeLuca} De Luca A., Molendi S. 2004, A\&A 419, 837
\bibitem[DeJager et al. 1996]{DeJager} De Jager O.C., Harding A.K.,
Strickman M.S. 1996, ApJ 460, 729
\bibitem[Dodson et al. 2003]{Dodson} Dodson R., Legge, D. et al. 2003,
ApJ 596, 1137
\bibitem[Fiore et al., 1999]{fiore} Fiore F., Guainazzi M., Grandi P.
1999 in Cookbook for {\it Beppo}SAX NFI Spectral Analysis, {\it Beppo}SAX Science
Data Center, version 1.2
\bibitem[Helfand et al. 2001]{Helfand} Helfand D.J., Gotthelf E.V.,
Halpern J.P. 2001, ApJ 556, 380
\bibitem[Kanbach et al. 1994]{Kanbach} Kanbach G., Arzoumanian Z. et al.
1994, A\&A 289, 855
\bibitem[kargalt 2003]{Kargaltsev1} Kargaltsev O.Y., Pavlov G. 2003, in
Young Neutron Stars and Their Environments, IAU Symp. 218,
F. Camilo and B.M. Gaensler eds., p. 195
\bibitem[KennellCoroniti 1984]{Kennell} Kennell C.F., Coroniti F.V. 1984, ApJ 283, 694
\bibitem[Kirsh et al 2004]{kirsh} Kirsh M.G.F, Altieri B. et al. 2004, 
{\it XMM Newton (cross) calibration}, XMM-SOC-CAL-TN-0055
\bibitem[Marty et al. 2002]{marty} Marty, P.B., Kneib, J.-P.et al. 2002,
Proc. SPIE. vol 4851, p. 208
\bibitem[Mignani et al., 2003]{mignani}Mignani R, De Luca A., Kargaltsev O. et al. 2003,
ApJ 594, 419 
\bibitem[Mori et al., 2003]{mori}Mori K., Hailey C.J. et al. 2004,
AdvSpRes 33, 503
\bibitem[Molendi  1998]{molendi}Molendi S., 1998 in The Active
X-ray Sky: Results from {\it Beppo}SAX and RXTE, ed. L. Scarsi, H. Bradt, P.
Giommi, \& F. Fiore; Nuclear Physics B, Proc. Suppl. 69, 563
\bibitem[Pavlov et al. 2001a]{Pavlova} Pavlov G.G., Kargaltsev O.Y.et al. 
2001a, ApJ 554, L189
\bibitem[Pavlov et al. 2001b]{Pavlovb} Pavlov G.G., Zavlin V.E. et al. 2001b,
ApJ 552, L129
\bibitem[Pavlov et al. 2003]{Pavlov3} Pavlov G.G., Teter M.A. et al. 2003,
ApJ 591, 1157
\bibitem[SefakoDeJager 2003]{Sefako} Sefako R.R., De Jager O.C. 2003, ApJ 593, 1013
\bibitem[Ulmer et al.1995]{ulmer95} Ulmer M.P., Matz S.M., Grabelsky D.A. et al. 1995,
ApJ 448, 356
\bibitem[Weiler Panagia, 1978]{weiler78} Weiler K.W., Panagia N. 1978,
A\&A 70, 419
\bibitem[Willmore et al.1992]{willmore92} Willmore A.P., Eyles C.J. et
al. 1992, MNRAS 254, 139
\end{thebibliography}
\end{document}